\documentclass[12pt]{article}

\usepackage{graphicx}
\usepackage{qcmc06}

\def\sH{\mathcal H}
\def\N#1{|\!|#1|\!|}
\def\sK{\mathcal K}
\def\Tr{\mathrm{Tr}}
\def\<{\langle}
\def\>{\rangle}
\def\Span{\mathrm{Span}}
\begin{document}

\pagestyle{empty}

\title{Optimal estimation of ensemble averages from a quantum
  measurement}

\author{Paolo Perinotti\refnote{1,2} and Giacomo M.
  D'Ariano\refnote{1}}

\affiliation{\affnote{1}Dipartimento di Fisica ``A. Volta'',\\
  via Bassi 6, 27100 Pavia, Italy\\
  \affnote{2}perinotti@fisicavolta.unipv.it}

\begin{abstract}
  We consider the general measurement scenario in which the ensemble
  average of an operator is determined via suitable data-processing of
  the outcomes of a quantum measurement described by a POVM. After
  reviewing the optimization of data processing that minimizes the
  statistical error of the estimation, we provide a compact formula
  for the evaluation of the estimation error.
\end{abstract}

\setcounter{section}{0}

\section{Introduction}

A measurement that can be performed in the lab is described by a POVM
(acronym for Positive Operator-Valued Measure), namely a set of
(generally nonorthogonal) positive operators $P_i\geq0$, $1\leq i\leq
N$ which resolve the identity $\sum_{i=1}^NP_i=I$ similarly to the
orthogonal projectors of an observable \cite{vonn}. The probability
distribution of the $i$th outcome is given by the Born rule
\begin{equation}\label{Born}
  p(i|\rho)=\Tr[P_i\rho]
\end{equation}
$\rho$ being the density operator of the state.  By such a measurement
one can experimentally determine the ensemble averages of (generally
complex) operators $X$. Clearly this is possible if $X$ can be
expanded over the POVM elements (mathematically we denote this
condition as $X\in\Span\{P_i\}_{i=1,N}$). This means that there exists
a set of coefficients $f_i[X]$ such that
\begin{equation}
  X=\sum_{i=1}^Nf_i[X]P_i,\quad X\in{\mathcal S}:=\Span\{P_i\}_{i=1,N}
  \label{expanX}
\end{equation}
When ${\mathcal S}\equiv\mathcal{B}(\sH)$ (i.~e. when all operators
can be expanded over the POVM), then the measurement is
informationally complete.  Obviously, once the expansion
(\ref{expanX}) is established one can obtain the ensemble average of
$X$ by the following averaging
\begin{equation}
\<X\>=\sum_{i=1}^Nf_i[X]p(i|\rho),
\end{equation}
where the probability distribution is given in Eq. (\ref{Born}).

The above general measurement procedure opens the problem of finding
the coefficients $f_i[X]$ in Eq. (\ref{expanX}), namely the
data-processing of the measurement outcomes needed to determine the
ensemble average of $X$. In general the coefficients $f_i[X]$ are not
unique (if $N>\dim(\mathcal{S})$), and one then wants to optimize the
data-processing according to a practical criterion, typically
minimizing the statistical error. This problem has been solved in the
general case in \cite{prl}, and its solution will be reviewed in this
paper. Here in addition We will present a simple formula for the
minimum estimation error for arbitrary operator $X$.

Notice that although the processing functions are intrinsically linear
in the definition (\ref{expanX}), there is no guarantee that the
optimal ones are linear in $X$. Remarkably, however, the optimal
processing function is indeed linear in $X$, and depends only on the
POVM and, in a {\em Bayesian scheme}, on the ensemble of possible
input states. The derivation of the optimal data-processing function
requires elementary notions of frame theory \cite{fram,banfram} and
linear algebra \cite{bhapat}, which will be introduced in the first
part of the paper.

\section{mathematical tools}

A frame in a Hilbert space $\sK$ is a set of vectors
$\{\phi_n\}\subseteq\sK$ satisfying the property
\begin{equation}
  a\N{\psi}^2\leq\sum_n|\<\phi_n|\psi\>|^2\leq b\N{\psi}^2,
\end{equation}
for all $\psi\in\sK$, with fixed $0<a\leq b<\infty$. The starting
theorem in frame theory states that the set $\{\phi_n\}$ is a frame
iff the positive operator, called {\em frame operator}
\begin{equation}
  F=\sum_n|\phi_n\>\<\phi_n|,
\end{equation}
is bounded and invertible. In this case we can define the {\em
  canonical dual frame} $\{\chi_n\}$ by the following formula
\begin{equation}
  |\chi_n\>=F^{-1}|\phi_n\>,
\end{equation}
and all the vectors $\psi\in\sK$ can be written as a linear
combination of the vectors $\{\phi_n\}$ as follows
\begin{equation}
  \label{exp}
  |\psi\>=\sum_n|\phi_n\>\<\chi_n|\psi\>.
\end{equation}
When the frame is made of linearly dependent vectors, the choice of
the coefficients in the expansion Eq.~(\ref{exp}) is not unique, and
all alternate choices are provided by {\em alternate dual frames}
$\{\eta_n\}$, classified by the relation \cite{li}
\begin{equation}
|\eta_n\>=|\chi_n\>+|\delta_n\>-\sum_m|\delta_m\>\<\phi_m|\chi_n\>,
\end{equation}
where $\{\delta_n\}\subseteq\sK$ is an arbitrary set of vectors. This
theorem is useful in our case because we can consider the POVM
elements $P_i$ as vectors in the space of Hilbert-Schmidt
operators---which for finite dimensional systems are all possible
operators $X$---and they provide a frame in the space $\mathcal S$.
Frame theory then solves the problem of finding all possible sets of
coefficients $f_i[X]$ in Eq.~(\ref{expanX}), which are simply given by
the scalar products $\<D_i|X\>:=\Tr[D_i^\dag X]$, $\{D_i\}$ being an
alternate dual for the frame $\{P_i\}\subseteq\mathcal S$.\par

In order to answer the main question of the paper, namely which is the
dual frame $\{D_i\}$ providing the minimum statistical error, we will
first show that the statistical error can be written in terms of a
norm for the vector $\{f_i[X]\}$ of coefficients. Indeed, if we
consider the ensemble of possible input states $\{\rho_k,p_k\}$, we
can define the statistical error in a fixed state $\rho_k$, and use
its average over the ensemble as a figure of merit. We have
\begin{equation}
  \label{noisx}
  \delta_D(X):=\sum_{i=1}^N\Tr[\rho_{\mathcal E}P_i]|f_i[X]|^2-\overline{\<X\>^2}_{\mathcal E},
\end{equation}
where $\rho_{\mathcal E}:=\sum_kp_k\rho_k$, and
$\overline{\<X\>^2}_{\mathcal E}:=\sum_kp_k|\Tr[\rho_k X]|^2$. The
second term in Eq.(\ref{noisx}) does not depend on the choice of the
dual, then the minimization problem can be stated as the minimization
of the norm
\begin{equation}
  \N{f[X]}_\pi:=\sum_{i=1}^N\pi_i|f_i[X]|^2,
\end{equation}
where $\pi_i=\Tr[P_i X]$. If we now consider the following linear map
that takes vectors of coefficients to operators
\begin{equation}
  \Lambda:c\to \sum_{i=1}^N c_iP_i,
\end{equation}
its matrix elements are given by $\Lambda_{mn,i}=(P_i)_{mn}$. One can
easily prove that all generalized inverses $\Gamma$ of $\Lambda$,
satisfying $\Lambda\Gamma\Lambda=\Lambda$, have matrix elements
$\Gamma_{i,mn}=(D_i^*)_{mn}$ where $\{D_i\}$ is an alternate dual
frame for $\{P_i\}$. The minimum noise can be obtained through the
{\em minimum norm} generalized inverse $\Gamma$ that must satisfy the
relation \cite{dps}
\begin{equation}
  \pi\Gamma\Lambda=\Lambda^\dag\Gamma^\dag\pi,
\end{equation}
where $\pi$ is the positive diagonal matrix with eigenvalues
$\pi_{i}$.


\section{minimization of error}
\label{sec-3}

Since the minimum norm generalized inverse is unique and does not
depend on the vector, the optimal dual does not depend on $X$, and the
function $f_i[X]=\<D_i|X\>$ is linear, as anticipated. One can prove
that the optimal dual frame $\{D_i\}$ corresponding to such $\Gamma$
is unique and can be expressed as follows \cite {prl}
\begin{equation}
  D_i=\Delta_i-\sum_j\{[(I-M)\pi(I-M)]^{\ddag}\pi\}_{ij}\Delta_j,
\end{equation}
where $\{\Delta_i\}$ is the canonical dual and $M$ is the projection
matrix with elements $M_{ij}=\Tr[\Delta_iP_j]$. The minimum noise for
$X$ can be expressed as
\begin{equation}
  \delta_D(X)=\<X|\Gamma^\dag\pi\Gamma|X\>-\overline{\<X\>^2}_{\mathcal E}=\<X|\left(\sum_{i=1}^N\Tr[\rho_{\mathcal E}P_i]|D_i\>\<D_i|\right)|X\>-\overline{\<X\>^2}_{\mathcal E}.
\end{equation}
On the other hand, one can prove the following identity
\begin{equation}
  \Gamma^\dag\pi\Gamma\Lambda\pi^{-1}\Lambda^\dag=\Gamma^\dag\Lambda^\dag\Gamma^\dag\Lambda^\dag=\Gamma^\dag\Lambda^\dag=\sum_{i=1}^N|D_i\>\<P_i|=I_{\mathcal S}.
\end{equation}
This implies that
$\Gamma^\dag\pi\Gamma=(\Lambda\pi^{-1}\Lambda^\dag)^{-1}$, and
finally, one can express the minimum noise in terms of the POVM and
the ensemble only, as follows
\begin{equation}
  \label{minnois}
  \delta_D(X)=\<X|(\Lambda\pi^{-1}\Lambda^\dag)^{-1}|X\>-\overline{\<X\>^2}_{\mathcal E}.
\end{equation}
The optimal dual has been obtained in a completely different framework
in \cite{scott} in the particular case when $\rho_{\mathcal E}=\frac
Id$, and the figure of merit considered therein is the Hilbert-Schmidt
distance between the estimated state and the true state.

\section{conclusion}

In this paper we reviewed the problem of estimation of ensemble
averages of operators by indirect measurements, through a fixed
measurement whose statistics is described by a POVM $\{P_i\}$. The
coefficients for the expansion of of an operator on the POVM elements
provide the processing functions, and their calculation is possible in
principle by using elementary results in frame theory. The difficult
problem is to decide which processing function is the best in order to
minimize the statistical error in the estimation of the ensemble
average. We restated this problem as the inversion of a rectangular
matrix with the constraint of minimum norm. We reviewed the general
solution derived in Ref. \cite{prl}, and we present a synthetic
formula for the evaluation of the minimum noise in terms of the POVM
elements and the input ensemble.

\ack P. P. thanks A. J. Scott for an interesting exchange of e-mails,
which inspired Eq. (\ref{minnois}). This work has been supported by
Ministero Italiano dell'Universit\`a e della Ricerca (MIUR) through
PRIN 2005. P.  P.  acknowledges financial support by EC under project
SECOQC (contract n. IST-2003-506813)

\end{document}